\documentclass[a4paper,11pt]{article}
\input{euclid.sty}

\pdfoutput=1 
             
\usepackage{jcappub} 
\usepackage{lineno}

\newcommand{\refEq}[1]{Eq.~\ref{#1}}
\newcommand{\refF}[1]{Fig.~\ref{#1}}

\newcommand{\refS}[1]{section~\ref{#1}}

\newcommand{\numberthis}{\addtocounter{equation}{1}\tag{\theequation}}

\usepackage[T1]{fontenc} 
\usepackage{rotating} 

\title{A $(D_\tau,D_x)$-manifold with $N$-correlators of $N_s$-objects}

\author[a,1]{Pierros Ntelis,}

\affiliation[a]{(Aix Marseille Univ, CNRS/IN2P3, CPPM, Marseille, France)}

\emailAdd{ntelis.pierros <at>  gmail <point> com}

\abstract{
Context.
In this paper, we describe a mathematical formalism for a  $(D_\tau,D_x)$-dimensional manifold with $N$-correlators of $N_s$ types  of targeted source objects, with cross correlations and contaminants. 

Methodology. In particular, we build this formalism using simple notions of mathematical physics, field theory, topology, algebra, statistics n-correlators and Fourier transform. 

Results.  We present and discuss the applicability of this formalism in the context of cosmological scales, i.e. from astronomical scales to quantum scales, for which we give some intuitive examples, explicitly, for standard  spacetime dimensions and extra dimensions. 

Conclusion. We conclude that this study can be used as a guide to analyse and build models for future cosmological and collider experiments. Furthermore, this study opens the road of extra dimension studies.
}

\keywords{cosmology; mathematical physics ; gravity ; quantum ; field theory ; large scale structure of Universe  }

\begin{document}
\maketitle
\date
\flushbottom

\section{Introduction}\label{Introduction}

Motivated by nature, modern cosmological theories such as field theories, the standard model of cosmology, its alternatives \cite{2023FoPh...53...29N,NtelisSaid2025a,NtelisSaid2025b,NtelisSaid2025c}, the inflationary paradigm \cite{PhysRevD.23.347,1982PhLB..108..389L,PhysRevLett.49.1110,starobinsky1982dynamics,PhysRevD.28.679,1992NuPhB.372..421L,1992PhLB..284..215L}, 
and primordial non-Gaussianity 
\cite{2000ApJ...541...10M,2001PhRvD..63f3002K}, as well as observational searches of these theories \cite{2020A&A...641A...6P,2011PhRvD..84h3509H,2008JCAP...08..031S,2008ApJ...677L..77M,2008PhRvD..77l3514D,2019JCAP...09..010C,2020A&A...641A...9P} and some of their systematic effects \cite{2007ApJ...660...62K,2016PASJ...68...12P,2016arXiv160608864W,2019ApJ...879...15A,2019ApJ...876...32G,2020arXiv201000047M,2020MNRAS.495.1340F}, scientific terminology for randomness has been an ongoing exploring subject in several scientific domains. A random field has been discussed in several application, and it has been excessively studied using N-point correlation functionals (NPCF). The simplest random fields are the so called Gaussian random fields, which present vanishing NPCF for N larger than three. However, higher than or equal to three NPCF have found successful applications in several scientific applications: molecular physics \cite{doi:10.1063/1.2883660}; material science \cite{BERRYMAN198886}; field theory \cite{Dotsenko1991THREEPOINTCF}; diffusive systems \cite{PhysRevE.48.800,PhysRevE.72.031108}; quantum field theory \cite{Peskin:1995ev}; computational physics \cite{2021MNRAS.tmp.2749P,2021arXiv210610278P}; cosmology \cite{2021MNRAS.tmp.2749P,2021arXiv210610278P,2001misk.conf...71M}. Furthermore this study expands the studies of extra dimensions, such as alternatives to compactifications \cite{1999PhRvL..83.4690R}. The importance of studying generalised correlations, in generalised manifolds is motivated because they quantify the energetic fluctuations of the universe, which are part of the actionic fluctuations \cite{2023FoPh...53...29N}. Note that we introduce a lot of the mathematical concepts used in this study, which renders this study a good pedagogical guideline for future studies.

These NPCFs try to quantify models of natural systems which most of them are built on manipulation of ingredients of the action principle (see \cite{2023FoPh...53...29N} and references therein). \citet{2021arXiv210610278P} have mapped the NPCF in D-dimensions, while \citet{2016PASJ...68...12P} have described a mathematical framestudy which is also applicable to Euclid \cite{2011arXiv1110.3193L} telescope for large scale structure (LSS) surveying, in which the line-misidentification is treated with modelling several contaminants of a targeted object selection, using only the 2PCF. A number of LSS experiments can benefit from our study, which deals with contaminants to their target sample of study, such as the Dark Energy Spectroscopy Instrument (DESI) \cite{2016arXiv161100036D}, Legacy Survey of Space and Time (LSST) \cite{2017ApJS..233...21R}, and Nancy Grace Roman Space Telescope \cite{2016PASJ...68...12P} and the Euclid satellite \cite{2022A&A...657A..91E}. Our study is also applicable to gravitational wave (GW) experiments, such as Virgo/LIGO experiments \cite{PhysRevLett.116.061102} and Einstein Telescope \cite{2020JCAP...03..050M}. Our study is also applicable to high energy physics experiments, such as the Large Hydron Collider (LHC)\cite{2019arXiv190304497A}, testing the standard model of particle physics, through elementary particle interactions, described by quantum field theory. 
In this study, we basically derive a short methodology and express equations describing N-point auto- and cross- correlation functionals in generalised dimensional spacetime manifold, denoted as ($D_\tau,D_x$)-dimensional manifold, of $N_{\rm s}$ types of objects, using fundamental mathematical principles and going beyond \citet{2016PASJ...68...12P}, and \citet{2021arXiv210610278P}, from the theoretical perspective. 

We present the applicability of such formalism to astronomical systems (the largest possible scales, i.e. astronomical scales, large scale structure) and to quantum systems (the smallest possible scales, i.e. quantum scales, small scale structure). Note that there are some different definitions of cosmological, astronomical and quantum scales in the literature. In this study, cosmological scales (CS) are the ones containing any scales appear in our cosmos, therefore it includes both quantum scales (QS) and astronomical scales (AS). Astronomical scales include scales between an astronomical unit (AU) to 
$8.6$ Gpc, which is the physical size of our universe. Note that an AU is equal to $149.6 \times 10^6$ kilo meters, which is basically the mean distance between the centre of earth to the centre of our solar system, the sun, while a pc is equal to $3.086 \times 10^{13}$ km. The quantum scales is the range of scales between the atomic scales, starting from millions of fempto meters (a fempto meter is equal to $10^{-15}$ m) to Planck scales, defined by $l_{\rm P} = 10^{-35}$ m, denoting the smallest possible scales. Note that these numbers can change assuming a different cosmology, away from the standard one.

Note that at AS the measurements are usually done in redshift space, which often is translated into distance measurements. However in order to study the time evolution of our universe, we usually bin large proportions of redshift measurements. Therefore, our model of course has to include any time variation. In this study, we extend the previous studied NPCF models, by including mutliple time variables and perturbed metrics. The multiple time variables is justified in the context of searches of extra dimensions. The perturbed metric is justified due to the fact that we considering cross correlations in our study, the one of photons from cosmic microwave background and galaxy densities, namely the ISW effect \cite{1967ApJ...147...73S}. Furthermore, we consider neutrino physics. Therefore our model should explicitly state the perturbations taken into account. These are studies that are well defined studied and performed, and therefore we include perturbations in our model.


Our study builds on the work of \citet{2021arXiv210610278P} by broadening the dimensional and applicative scope of N-point correlation functions (NPCFs). While  \citet{2021arXiv210610278P} concentrate on D-dimensional correlators specifically tailored for large-scale structure (LSS) analyses, our investigation extends this framework to a $(D_\tau, D_x)$-dimensional manifold, incorporating multiple temporal $(D_\tau)$ and spatial $(D_x)$ dimensions to accommodate a more versatile spacetime representation. Moreover, their focus is on observables of targeted objects within LSS, whereas we develop a generalized observable formalism applicable across both LSS (astronomical scales) and small-scale structure (quantum scales), enhancing its utility for diverse scientific inquiries. Additionally, \citet{2021arXiv210610278P} do not consider contaminants or temporal dynamics, whereas our approach includes the effects of contaminants on targeted objects and examines their growth over time, enabling the formalism to address realistic observational challenges in experiments ranging from DESI to LHC.

In summary, this study discusses the formalism of the generalised manifold concept of the problem of tracer and contamination of NPCF observables of current and future cosmological experiments. For a generalisation of this manifold metric pair please read the companion paper \cite{Ntelis2025_AMMP} and an application of the companion paper please read \cite{Aprobabilisticexpandinguniverse}. Another generalisation that can be applied to this study is the one from advancing tensor theories \cite{Ntelis2025_ATT}, and category theories \cite{Advancing_categories_with_functors_of_functors}.
In the future, this formalism may be applied and encoded in manifold learning software systems which is a type of machine learning system applied on manifold information as was successfully used in \citet{2021ApJ...912...71B}.

\section{Methodology}
In this section, we present the methodology followed to accomplish our study. We start by discussing several concepts in generalised manifolds and generalised correlators of cosmic objects. Then, we present the equations for N-point correlators of an object in a ($D_\tau,D_x)$-manifolds, including and going beyond the $(1,3)$- dimensions. Then we present the equations of a combination of a variety of objects, and then with some distortions from contaminants.

\subsection{Generalised manifolds and correlators of cosmic objects}\label{sec:norder_Correlations_of_tracers}

In nature, we consider some simple (1,3)-manifolds in order to explain the physical phenomena. Be it from astronomical scales to cosmological scales most of the natural physical systems are explained with differential equations composed by one temporal component and three spatial ones. Furthermore, the NPCF is considered a tool which can be used to analyse several problems in nature. In this study, we expand and generalise further these concepts.

\subsubsection{A $(D_\tau,D_x)$-dimensional manifold}

Let's consider a general $D$-dimensional manifold, $\mathcal{M}^{D}$, where $D$ denotes the dimensions of the manifold. The tensor product is denoted with $\otimes$. Let's consider that there is a submanifold which has a dimensional set which can be denoted with $(D_\tau,D_x) \subset D$,  where $D_\tau$ denotes the number of dimensions of conformal times, and $D_x$ denotes the number of dimensions of spatial spaces. Then each $D$-tuplet $\vec{\tau x} \equiv  \left\{ \vec{\tau}, \vec{x} \right\}$ denotes a point of the manifold which is written by $\mathcal{M}^{D}\supset \mathcal{M}^{(D_\tau,D_x)} = \mathcal{M}^{D_\tau} \otimes \mathcal{M}^{D_x}$. For a generalisation of this manifold metric pair please read \cite{Ntelis2025_AMMP}.

We can construct an arbitrarily large number of line elements for this particular manifold, which is only limited by our imagination and experiments. However, here we are going to consider the following line element. The line element of an expanding, perturbed, homogeneous, isotropic, Anti-de-Sitter $(D_\tau,D_x)$-manifold can be given by
\begin{align}\label{eq:generalisedAdS}
	ds_{(D_{\tau},D_x)}^2 
	&= g^{(D_{\tau},D_x)}_{\alpha\beta} (x) dx^\alpha dx^\beta 
	\\ 
	&= a^2( \vec{\tau} ) \left[  e^{-2\Phi(\vec{\tau},\vec{x})} \sum_{i=1}^{D_x} \sum_{j=1}^{D_x} dx^i dx^j \delta_{ij} - e^{2\Psi(\vec{\tau},\vec{x})} \sum_{b=1}^{D_\tau} (d\tau_b)^2 \right]\; ,
\end{align}
where $dx^\alpha=\left\{ d\tau_1, \dots, d\tau_{D_\tau}, dx_1, \dots, dx_{D_x} \right\}$ denotes the infinitesimal element in a generalised Minkowski $(D_\tau,D_x)$-manifold, $g^{(D_{\tau},D_x)}_{\alpha\beta} (x)$ is the accompany metric tensor, $a(\vec{\tau})$ is the scale factor defined in the $\mathcal{M}^{(D_\tau,D_x)}$-manifold, while $e^{2\Psi(\vec{\tau},\vec{x})}\simeq 1+2\Psi(\vec{\tau},\vec{x})$ and $e^{2\Phi(\vec{\tau},\vec{x})}\simeq 1+2\Phi(\vec{\tau},\vec{x})$ describe the perturbations of the metric. Note that this is a generalised Anti-de Sitter spacetime and for $(D_\tau,D_x)=(1,3)$ this line element is reduced to the standard generalised Minkowski spacetime which describes an expanding perturbed spacetime (EPST), i.e. 
\begin{align}
	ds^2_{\rm EPST} = a^2(\vec{\tau}) \left[ - e^{2\Psi(\vec{\tau},\vec{r})} d\tau^2 + e^{-2\Phi(\vec{\tau},\vec{r})} \sum_{i=1}^{3} \sum_{j=1}^{3} dx^i dx^j \delta_{ij}\right]	\; ,
\end{align} 
see \citet{1995ApJ...455....7M}. This line element described by \refEq{eq:generalisedAdS} can be transformed into $D_r$-spherical coordinates, as follows. This means that the spatial component can be considered using $D_r$-spherical coordinate, while the temporal component remains unchanged. Therefore, we get 
\begin{align}
	ds^2_{(D_\tau,D_r)} = a^2(\vec{\tau}) \left[ - e^{2\Psi(\vec{\tau},\vec{r})}\sum_{b=1}^{D_\tau} (d\tau_b)^2 + e^{-2\Phi(\vec{\tau},\vec{r})} dr_{D_r}^2 \right] \; .
\end{align}
By considering different types of topological spaces
\footnote{ 
Note that while curvature is a key distinguishing feature in the Friedmann-Lemaître-Robertson-Walker models—open (\(k < 0\), hyperbolic), flat (\(k = 0\), Euclidean), and closed (\(k > 0\), spherical)—these spaces are also topologically distinct. Topology extends beyond the number of holes (e.g., genus) to properties like compactness and connectivity. Open and flat universes, modeled by hyperbolic manifolds or \(\mathbb{R}^3\), are non-compact and infinite, while closed universes, like the 3-sphere, are compact and finite. For instance, the hyperbolic plane (open) and \(\mathbb{R}^2\) (flat) are not homeomorphic to the 2-sphere (closed) due to differences in compactness and fundamental groups, even if they share the same number of holes (none). These topological distinctions, alongside curvature, influence cosmological observables and the universe's expansion dynamics. 
}
, i.e. closed ($k>0$), flat ($k=0$) and open($k<0$), we define $dr_{D_r}^2 $ line element as follows,
\begin{align}
	dr_{D_r}^2 = dr^2 + S_{k}^2(r)d\Omega_{D_r-1}^2
\end{align}
where 
\begin{align}
	S_{k}(r) = 
	\left\{
	\begin{matrix}
		|k|^{-1/2} \sin(r\sqrt{k}) &, k > 0 \\
		r \hspace{1cm} &, k=0 \\
		|k|^{-1/2} \sinh(r\sqrt{k}) &, k < 0	\\	
	\end{matrix}
	\right.
\end{align}
while we have that 
\begin{align}
	d \Omega_{D_r-1}^2 = d\theta_i d\theta_j g_{ij}^{(D_r-1)} 
\end{align}
where 
\begin{align}
	g_{ij}^{(D_r-1)}  = 
	\left( 
	\begin{matrix}
	1 & 0                        & 0 & \dots & 0 & 0 \\
	0 & \sin^2\theta_{1} & 0 & \dots & 0 & 0 \\
	0 & 0 & \sin^2\theta_{1} \sin^2\theta_{2} & \dots & 0 & 0 \\
	0 & 0 &  0  & \dots & 0 & 0 \\
	. & . & . &  \dots  &  . & . \\
	. & . & . &  \dots  & . & . \\
	. & . & . &  \dots  & . & . \\
	0 & 0 & 0 &  \dots & 0 & \prod_{i=1}^{D_r-1} \sin^2\theta_i 	
	\end{matrix}
	\right)
\end{align}
while 
\begin{align}
	 d\theta_i \in \left\{ d\theta_1, \dots, d\theta_{D_r-1} \right\}
\end{align}
where $r\in \mathbb{R}^{+}$, 
	  $\theta_{i\in[1,D_r-2]} \in \left[0,\pi\right]$, 
and    $\theta_{D_r-1} \in \left[0,2\pi\right]$. Note that in many applications, we can use interchangeably the $D_x$-cartesian coordinates and the $D_r$-spherical coordinates.

\subsubsection{$N$-point correlators of an object in a $(D_\tau,D_x)$-manifold}

Developing further the study from \cite{2021arXiv210610278P}, let's build an observed quantity of $O$ objects, as follows. 
Consider a $(D_\tau, D_x)$-manifold (e.g. generalised Minkowski manifold) with an associated metric and observable quantity. This observable quantity is denoted with a complex-valued random field functional, $O$: $\mathcal{M}^{(D_\tau, D_x)} \rightarrow \mathbb{C}$, where $\mathbb{C}$ is the complex number set. Then, the NPCF $F^{(N)}$: $\mathcal{M}^{(D_\tau, D_x)}\otimes \dots \otimes \mathcal{M}^{(D_\tau, D_x)} \rightarrow \mathbb{C}$, is formally defined as: 
\begin{align}
F^{(N)}(\vec{\tau}, \vec{x}_1, \dots, \vec{x}_{N-1};\vec{s}) &\equiv \mathbb{E}_O \left[ O(\vec{\tau}, \vec{s}) \cdot{} O(\vec{\tau}, \vec{s}+\vec{x}_1) \cdot{} {\dots} \cdot{}  O(\vec{\tau},\vec{s}+\vec{x}_{N-1})  \right] \; ,
\end{align}
where $\mathbb{E}$ represent the statistical average over realisation of an $O$ object, while 
$\vec{s}$ and $\vec{x}_i$ are absolute and relative positions on the manifold $\mathcal{M}^{D_x}$. Note that $\vec{\tau}$ defines a temporal position to the manifold $\mathcal{M}^{D_\tau}$ and we have assumed $N\geq 2$. In the case which the random field is statistically homogeneous, all correlators must be independent of the
absolute position $s$. This leads to the popular NPCF estimator given by 
\begin{align}\label{eq:NPCF}
\hat{F}^{(N)}(\vec{\tau}, \vec{x}_1, \dots, \vec{x}_{N-1}) &\equiv \langle O (\vec{\tau}, \vec{s}) \cdot{} O (\vec{\tau}, \vec{s}+\vec{x}_1) \cdot{} {\dots} \cdot{} O (\vec{\tau},\vec{s}+\vec{x}_{N-1})  \rangle_s \; ,
\end{align}
where $\langle \dots \rangle_s \equiv V_{D_x}^{-1} \int_{\mathcal{M}^{D_x}} d^{D_x} \vec{s}\; \left[ \dots \right] $ is the volume average integration over the $D_x$-dimensional spatial volume $V_{D_x}$ contained in the $\mathcal{M}^{(D_\tau,D_x)}$-manifold, assuming the ergodic theorem in the $\mathcal{M}^{D_x}$-manifold. Note that the NPCF depends only on $(N-1)$ positions. 
In spatial Fourier space
the observable is 
\begin{align}
\tilde{O}(\vec{\tau},\vec{k}) \equiv \int_{\mathcal{M}^{D_x}} d^{D_x} \vec{s}\; e^{-i\vec{k}\cdot{\vec{s}}} O(\vec{\tau},\vec{s}) 
\end{align}
while, using a Fourier Transform (FT), the N-order correlation function in Fourier space,  is
\begin{equation}\label{eq:NPCFF}
\hat{\tilde{F}}^{(N)}(\vec{\tau}, \vec{k}_1, \dots, \vec{ k}_{N-1}) = \langle \tilde{O} (\vec{\tau}, \vec{q}+\vec{k}_1) \cdot{} {\dots} \cdot{}  \tilde{O} (\vec{\tau},\vec{q}+\vec{k}_{N-1})  \rangle_q \; ,
\end{equation}
where $\langle \dots \rangle_q \equiv V_{D_q}^{-1} \int_{\mathcal{M}^{D_q}} d^3 \vec{q} $ is the averaged integration of the observable. Note that we will simplify the rest of the discussion and we are going simply the notation and we will not use $\hat{}$ symbol to denote an estimator and we will not use the $\tilde{}$ symbol to denote an FT, since it will be clear from the context.

\subsubsection{A combination of a variety of objects}
In several application, it has been demonstrated that one can have a variety of different types of source objects that can be targeted from a set of targets, $S_s$, which can be correlated in a particular field configuration. Along these lines, we can define $N_{\rm s}$ types of objects which exist in the same $\mathcal{M}^{(D_{\tau},D_x)}$ manifold. This means that the total observed objects will be the sum of such $N_{\rm s}$ objects denoted with the observables of $O_t$, where $t$ denotes the type of the object, and we write
\begin{align}\label{eq:N_s_objects}
		O(\vec{\tau}, \vec{x}) 
		\equiv \sum_{s=1}^{N_{\rm s}} O_t(\vec{\tau}, \vec{x}) 
		 \; .
\end{align}
Note that we can decompose the targeted observable as 
\begin{align}
	O_{s}(\vec{\tau},\vec{x})= 	O_{\rm u}(\vec{\tau}_i,\vec{x}) D_s(\vec{\tau},\vec{x})
	\; ,
\end{align}
where $O_{\rm u}(\vec{\tau}_i,\vec{x})$ is a universal observable which depend in some initial temporal space of $D_\tau$-dimensions denoted with $\vec{\tau}_i$, while $D_s (\vec{\tau},\vec{x})$ encapsulate the rest $(D_\tau,D_x)$ spacetime dependence. Then the observable becomes 
\begin{align}\label{eq:N_s_objects_decomposition_1}
	O(\vec{\tau}, \vec{x}) 
		\equiv O_{\rm u}(\vec{\tau}_i,\vec{x})  \sum_{s=1}^{N_{\rm s}} D_s(\vec{\tau},\vec{x}) \; ,
\end{align}

Note that, by substituting \refEq{eq:N_s_objects} and/or \refEq{eq:N_s_objects_decomposition_1} to Eqs. \ref{eq:NPCF} and \ref{eq:NPCFF}, we can calculate the auto- and cross- NPCF and its Fourier transform of $N_{\rm s}$ types of objects.

By substituting \refEq{eq:N_s_objects_decomposition_1} to Eqs. \ref{eq:NPCF}, we get
\begin{align}\label{eq:NPCF_variety_of_objects_decomposed}
\small
	F^{(N)}(\vec{\tau}, \vec{x}_1, \dots, \vec{x}_{N-1}) &\equiv F^{(N)}_{\rm u} (\vec{\tau}_i, \vec{x}_1, \dots, \vec{x}_{N-1}) 
	\cdot{} \nonumber\\
	&\cdot{}
	\langle  \sum_{s=1}^{N_{\rm s}} D_s (\vec{\tau}, \vec{s}) \cdot{} D_s (\vec{\tau}, \vec{s}+\vec{x}_1) \cdot{} {\dots} \cdot{} D_s (\vec{\tau},\vec{s}+\vec{x}_{N-1})  \rangle_s \; ,
\end{align}
where 
\begin{align}
	F^{(N)}_{\rm u} (\vec{\tau}_i, \vec{x}_1, \dots, \vec{x}_{N-1}) \equiv 
	\langle O (\vec{\tau}_i, \vec{s}) \cdot{} O (\vec{\tau}_i, \vec{s}+\vec{x}_1) \cdot{} {\dots} \cdot{} O (\vec{\tau}_i,\vec{s}+\vec{x}_{N-1})  \rangle_s \; .
\end{align}
In case that every decomposition factor depends only on time, $D_s (\vec{\tau},\vec{x}) \rightarrow D_s (\vec{\tau})$, then we get simply:
\begin{align}
	F^{(N)}_{\rm simplified, 1} (\vec{\tau}, \vec{x}_1, \dots, \vec{x}_{N-1}) &\equiv F^{(N)}_{\rm u} (\vec{\tau}_i, \vec{x}_1, \dots, \vec{x}_{N-1})  \left[ \sum_{s=1}^{N_{\rm s}} D_s(\vec{\tau}) \right]^N \; ,
\end{align}

While in Fourier space we get simply in this case:
\begin{align}\label{eq:NPCF_decomposition_simplified}
	F^{(N)}_{\rm simplified, 1} (\vec{\tau}, \vec{k}_1, \dots, \vec{k}_{N-1}) &\equiv F^{(N)}_{\rm u} (\vec{\tau}_i, \vec{k}_1, \dots, \vec{k}_{N-1})  \left[ \sum_{s=1}^{N_{\rm s}} D_s(\vec{\tau}) \right]^N \; ,
\end{align}
where 
\begin{align}
	F^{(N)}_{\rm u} (\vec{\tau}_i, \vec{k}_1, \dots, \vec{k}_{N-1}) \equiv 
	\langle O (\vec{\tau}_i, \vec{s}) \cdot{} O (\vec{\tau}_i, \vec{q}+\vec{k}_1) \cdot{} {\dots} \cdot{} O (\vec{\tau}_i,\vec{q}+\vec{k}_{N-1})  \rangle_q \; .
\end{align}
This means that in some special configurations, in which an observable can be decomposed into an observable that depends on some initial time and spatial space, while the growth of each tracer depends only on time, the NPCF of this decomposed observable can be simplified in simple functional form, which is given by \refEq{eq:NPCF_decomposition_simplified}. 
Note that we can decompose the targeted observable as 
\begin{align}
	O_{s}(\vec{\tau},\vec{x})= 	O_{\rm u}(\vec{\tau}_i,\vec{x}_b) D_s(\vec{\tau},\vec{x})
	\; ,
\end{align}
where $O_{\rm u}(\vec{\tau}_i,\vec{x}_b)$ is a universal observable which depend in some initial temporal space of $D_\tau$-dimensions denoted with $\vec{\tau}_i$, while a boundary space of $D_x$ dimensions denoted by $\vec{x}_b$ while  $D_s (\vec{\tau}_i,\vec{x})$ encapsulate the $(D_\tau,D_x)$ generalised spacetime dependence. Then the observable becomes 
\begin{align}\label{eq:N_s_objects_decomposition_2}
	O(\vec{\tau}, \vec{x}) 
		\equiv O_{\rm u}(\vec{\tau}_i,\vec{x}_b)  \sum_{s=1}^{N_{\rm s}} D_s(\vec{\tau},\vec{x}) \; ,
\end{align}

Note that, by substituting \refEq{eq:N_s_objects_decomposition_2} to Eqs. \ref{eq:NPCF} and \ref{eq:NPCFF}, we can calculate the auto- and cross- NPCF and its Fourier transform of $N_{\rm s}$ types of objects.

By substituting \refEq{eq:N_s_objects_decomposition_2} to Eqs. \ref{eq:NPCF}, and some rearranging, we get
\begin{align}\label{eq:NPCF_variety_of_objects_decomposed_2}
	F^{(N)}(\vec{\tau}, \vec{x}_1, \dots, \vec{x}_{N-1}) &\equiv O^N_{\rm u}(\vec{\tau}_i,\vec{x}_b)  \sum_{s=1}^{N_{\rm s}}  \langle D_s (\vec{\tau}, \vec{s}) \cdot{} D_s (\vec{\tau}, \vec{s}+\vec{x}_1) \cdot{} {\dots} \cdot{} D_s (\vec{\tau},\vec{s}+\vec{x}_{N-1})  \rangle_s \; ,
\end{align}
In case that every decomposition factor depends only on time, $D_s (\vec{\tau},\vec{x}) \rightarrow D_s (\vec{\tau})$, then we get simply:
\begin{align}
	F^{(N)}_{\rm simplified, 2} (\vec{\tau}, \vec{x}_1, \dots, \vec{x}_{N-1}) &\equiv O^N_{\rm u}(\vec{\tau}_i,\vec{x}_b)  \left[ \sum_{s=1}^{N_{\rm s}} D_s(\vec{\tau}) \right]^N \; ,
\end{align}

\subsubsection{Distortion from contaminants}\label{sec:distortion_from_contaminants}
In case we would like to distinguish a targeted object category, in respect of several others which contaminate the targeted object category we can think the following, after inspired by \cite{2016PASJ...68...12P}. We can have the targeted objects, and the contaminant objects. the targets belong to the  ($D_\tau,D_x$)-dimensional-T manifold, $\mathcal{M}^{(D_\tau,D_x)}_T$, while the contaminants belong to the  ($D_\tau,D_x$)-dimensional-C manifold, $\mathcal{M}^{(D_\tau,D_x)}_C$. This means that the observed quantity will be a combination of the $N_{\rm s}$ targeted objects denoted with the tensor, $O_{s}(\vec{\tau},\vec{x})$ and $N_{{\rm c}t}$ objects, i.e. contaminants of each target, belonging to the set $S_{cs}$, which contaminate each targeted object denoted with the tensor, $O_{cs}(\vec{\tau},\vec{\gamma}\hspace{1mm}^{-1}\cdot{}\vec{x})$, where $\vec{\gamma}$ is a distortion factor tensor, which can be defined differently for each application. 
We assume that in general this distortion factor tensor is due to the fact that the contaminants are coming to the targeted manifold, from the $\mathcal{M}^{(D_\tau,D_x)}_C$. Therefore, it has a $(D_\tau,D_x)$ dependence and it can be denoted by the tensor $\vec{\gamma} \equiv \vec{\gamma_{tc}}(\vec{\tau},\vec{x})$. This can be achieved according to a factor of contamination of each target denoted with the tensor, 
	\begin{align}
	f_{cs}(\vec{\tau},\vec{x}) =  N_{cs}(\vec{\tau},\vec{x})/N_{\rm O} \; , 
	\end{align}	
where 
	\begin{align}
N_{\rm O} &= N_{\rm s} + N_{\rm cs} \\
&= \left( V_{ D_x} V_{D_\tau}\right)^{-1} \int_{V_{ D_x}} d^{D_x} \vec{x} \int_{V_{D_\tau}} d^{D_\tau} \vec{\tau} \sum_{s=1}^{N_{\rm s}} \left[ N_{t}(\vec{\tau},\vec{x}) +\sum_{c=1}^{N_{{\rm c} s}}N_{c s}(\vec{\tau},\vec{x}) \right] \; . 
	\end{align}
Therefore, the observed quantity is re-written as
\begin{align}\label{eq:Observed_distorted}
	O(\vec{\tau}, \vec{x}) 
	= \sum_{s=1}^{N_{\rm s}}  \left\{  \left[1- \sum_{c=1}^{N_{{\rm c} s}}f_{cs}(\vec{\tau},\vec{x}) \right]\; O_{s}(\vec{\tau},\vec{x})+ \sum_{c=1}^{N_{{\rm c} s}} f_{cs} (\vec{\tau},\vec{x})\; O_{c t}\left[\vec{\tau}, \vec{\gamma_{cs}}\hspace{1mm}^{-1}\left( \vec{\tau},\vec{x}\right)\cdot{} \vec{x}\right] \right\}
	\; .
\end{align}
Note that the distortion of the space component happens as 
\begin{align}
	\vec{y} =  \vec{\gamma_{cs}}\hspace{1mm}^{-1} \cdot{} \vec{x}
\end{align}

We also remind that FT implies:
 \begin{align}
 	\vec{x}_j &\rightarrow \vec{k}_j \\ 
	\vec{y}_j &\rightarrow \vec{q}_j = \vec{\gamma_{cs}}  \cdot{} \vec{k}_j 
\end{align}
which means that we can use the relations
 \begin{align}
 	d^{D_{x}} \vec{y} &= dy_{1} \cdot{} {\dots} \cdot{} dy_{D_x} = \gamma_{cs 1}^{-1} dx_{1}  \cdot{} {\dots} \cdot{}  \gamma_{cs D_x}^{-1} dx_{D_x} = d^{D_x}\vec{x} \prod_{d_x=1}^{D_x} \gamma_{cs d_x}^{-1}  \\
 	d^{D_{x}} \vec{q} &= dq_{1} \cdot{} {\dots} \cdot{} dq_{D_x} = \gamma_{cs 1} dk_{1}  \cdot{} {\dots} \cdot{}  \gamma_{cs D_x} dk_{D_x} = d^{D_x}\vec{k} \prod_{d_x=1}^{D_x} \gamma_{cs d_x}  
 \end{align}
 Now we can use the following relations:
\begin{align}
	O_{cs}(\vec{\tau},\vec{k}) &= \int d^{D_x} \vec{x} e^{-i\vec{k}\cdot{} \vec{x}} O_{cs} (\vec{\tau},\vec{y}) \\
	O_{cs}(\vec{\tau},\vec{k}) &= \prod_{d_x=1}^{D_x} |\gamma_{cs d_x}| \int d^{D_x} \vec{y} e^{-i\vec{q}\cdot{} \vec{y}}O_{cs}(\vec{\tau},\vec{y})\\
	&= \prod_{d_x=1}^{D_x} |\gamma_{cs d_x}| O_{cs}(\vec{\tau},\vec{q})
\end{align}
which means:
\begin{equation}
O_{cs}(\vec{\tau},\vec{k}) =\prod_{d_x=1}^{D_x} |\gamma_{cs d_x}| O_{cs}  (\vec{\tau},\vec{q}) \; ,
\end{equation} 
as well as the fact that 
\begin{align}
	O_{cs} (\vec{\tau},\vec{x}) 
	&= \int d^{D_x} \vec{k} e^{i\vec{k}\cdot{} \vec{x}} O_{cs} (\vec{x},\vec{q}) \\
	&= \left( \prod_{d_x=1}^{D_x} |\gamma_{cs d_x}| \right)^{-1} \int d^{D_x} \vec{q} 
	e^{ i \vec{q}\cdot{} \vec{y} } O_{cs} (\vec{x},\vec{q}) \\
	O_{cs} (\vec{\tau},\vec{x})  
	&= \left( \prod_{d_x=1}^{D_x} |\gamma_{cs d_x}| \right)^{-1} O_{cs} (\vec{\tau},\vec{y})  
\end{align}
which means:
\begin{equation}
O_{cs} (\vec{\tau}, \vec{y}) =  \prod_{d_x=1}^{D_x} | \gamma_{cs d_x}| O_{cs} (\vec{\tau},\vec{x})  \; .
\end{equation}
In the case which the distortion is the same for all dimensions and has only time dependence, $\vec{\tau}$, we have that 
\begin{align}
	\prod_{d_x=1}^{D_x} |\gamma_{cs d_x} (\vec{\tau}) | = |\gamma_{cs} (\vec{\tau})|^{D_x}
\end{align}
This means that the observable, which is distorted from contaminants, i.e. \refEq{eq:Observed_distorted}, becomes
\begin{align}\label{eq:Observed_distorted_step2}
	O(\vec{\tau}, \vec{x}) 
	= \sum_{s=1}^{N_{\rm s}} \left\{  \left[1- \sum_{c=1}^{N_{{\rm c} s}} f_{cs}(\vec{\tau},\vec{x}) \right] O_{s}(\vec{\tau},\vec{x})+ 
	\sum_{c=1}^{N_{{\rm c} s}} |\gamma_{cs} (\vec{\tau}) |^{D_x}f_{cs} (\vec{\tau},\vec{x}) O_{c s}(\vec{\tau},  \vec{x}) \right\}
	\; .
\end{align}


Note that we can also decompose the targeted observable as 
\begin{align}
	O_{s}(\vec{\tau},\vec{x})= 	O_{\rm u}(\vec{\tau}_i,\vec{x}) D_s(\vec{\tau},\vec{x})
	\; ,
\end{align}
where $O_{\rm u}(\vec{\tau}_i,\vec{x})$ is a universal observable which depend in some initial temporal space of $D_\tau$-dimensions denoted with $\vec{\tau}_i$, and the spatial space, $\vec{x}$ in $D_x$ dimensions, while $D_s (\vec{\tau}_i,\vec{x})$ encapsulate the rest $(D_\tau,D_x)$ generalised spacetime dependence. Note that also there exist a decomposition factor for the targeted contaminated observable defined as
\begin{align}
	O_{cs}(\vec{\tau},\vec{x})= O_{\rm u}(\vec{\tau}_i,\vec{x}) D_{cs}(\vec{\tau},\vec{x})
	\; .
\end{align}
Note that there is also a decomposition factor, of the targeted contaminated observable denoted with denoted with ${D^{\rm (F)}}_{cs}(\vec{\tau},\vec{x})$, and it is defined as
\begin{align}
	D_{cs}(\vec{\tau},\vec{x}) = D_s (\vec{\tau},\vec{x}) {D^{\rm (F)}}_{cs} (\vec{\tau},\vec{x})
\end{align}  
Note also that with the aforementioned decomposition, the \refEq{eq:Observed_distorted_step2} is analysed to 
\begin{align}
	O(\vec{\tau}, \vec{x}) 
	= O_{\rm u}(\vec{\tau}_i,\vec{x}) \sum_{s=1}^{N_{\rm s}} D_s(\vec{\tau},\vec{x}) \left\{  \left[1- \sum_{c=1}^{N_{{\rm c} s}} f_{cs}(\vec{\tau},\vec{x}) \right] + 
	\sum_{c=1}^{N_{{\rm c} s}} |\gamma_{cs}  (\vec{\tau}) |^{D_x} f_{cs} (\vec{\tau},\vec{x}) {D^{\rm (F)}}_{cs}(\vec{\tau},  \vec{x}) \right\}
	\; .
\end{align}
This means that we can define the decomposition factor of contamination and decomposition in spacetime functional, namely $\mathcal{DF}(\vec{\tau x})$, as 
\begin{align}
	\mathcal{DF}(\vec{\tau x}) \equiv \sum_{s=1}^{N_{\rm s}} D_s(\vec{\tau},\vec{x}) \left\{  \left[1- \sum_{c=1}^{N_{{\rm c} s}} f_{cs}(\vec{\tau},\vec{x}) \right] + 
	\sum_{c=1}^{N_{{\rm c} s}} |\gamma_{cs}  (\vec{\tau}) |^{D_x} f_{cs} (\vec{\tau},\vec{x}) {D^{\rm (F)}}_{cs}(\vec{\tau},  \vec{x}) \right\}
	\; .
\end{align}
This means that the observable can be defined as 
\begin{align}\label{eq:N_s_objects_N_c_t_contaminants_general_space}
	O(\vec{\tau}, \vec{x}) 
	= O_{\rm u}(\vec{\tau}_i,\vec{x}) \mathcal{DF}(\vec{\tau x}) 
	\; .
\end{align}
Now that by substituting \refEq{eq:N_s_objects_N_c_t_contaminants_general_space} to Eqs. \ref{eq:NPCF} and \ref{eq:NPCFF}, we can calculate the auto- and cross- NPCF and its Fourier transform of $N_{\rm s}$ types of objects, with $N_{{\rm c} s}$ types of contaminants.
In this case, by substituting \refEq{eq:N_s_objects_N_c_t_contaminants_general_space} to Eqs. \ref{eq:NPCF}, we get
\begin{align}
	F^{(N)}_{\rm simplified, 1}(\vec{\tau}, \vec{x}, \dots, \vec{x}_{N-1}) &= 	F^{(N)}_{\rm u}(\vec{\tau}_i, \vec{x}, \dots, \vec{x}_{N-1}) \cdot{} \nonumber \\
	& \langle \mathcal{DF}(\vec{\tau},\vec{s}) \cdot{} \mathcal{DF}(\vec{\tau}, \vec{x}+\vec{s}) \cdot{} \dots{} \cdot{} \mathcal{DF}(\vec{\tau}, \vec{x}_{N-1}+\vec{s}) \rangle_s \; ,
\end{align}
where 
\begin{align}
	F^{(N)}_{\rm u}(\vec{\tau}_i, \vec{x}, \dots, \vec{x}_{N-1}) \equiv \langle O_{\rm u}(\vec{\tau}_i,\vec{s}) \cdot{} 	O_{\rm u}(\vec{\tau}_i, \vec{x}+\vec{s}) \cdot{} \dots{} \cdot{} O_{\rm u}(\vec{\tau}_i, \vec{x}_{N-1}+\vec{s}) \rangle_s \; .
\end{align}
In the case where the decomposition function depends only on time $\mathcal{DF}(\vec{\tau},\vec{x}) \rightarrow \mathcal{DF}(\vec{\tau})$, we simply get
\begin{align}	
	F^{(N)}_{\rm simplified, 2}(\vec{\tau}, \vec{x}, \dots, \vec{x}_{N-1}) = F^{(N)}_{\rm u}(\vec{\tau}_i, \vec{x}, \dots, \vec{x}_{N-1}) \mathcal{DF}^N(\vec{\tau}) \; .
\end{align}


Note that we can also decompose the targeted observable as 
\begin{align}
	O_{s}(\vec{\tau},\vec{x})= 	O_{\rm u}(\vec{\tau}_i,\vec{x}_b) D_s(\vec{\tau},\vec{x})
	\; ,
\end{align}
where $O_{\rm u}(\vec{\tau}_i,\vec{x}_b)$ is a universal observable which depend in some initial temporal space of $D_\tau$-dimensions denoted with $\vec{\tau}_i$, while a boundary space of $D_x$ dimensions denoted by $\vec{x}_b$ while $D_s (\vec{\tau},\vec{x})$ encapsulate the $(D_\tau,D_x)$ generalised spacetime dependence. Note that also there exist a decomposition factor for the targeted contaminated observable defined as
\begin{align}
	O_{cs}(\vec{\tau},\vec{x})= O_{\rm u}(\vec{\tau}_i,\vec{x}_b) D_{cs}(\vec{\tau},\vec{x})
	\; .
\end{align}
Note that there is also a decomposition factor, of the targeted contaminated observable denoted with denoted with ${D^{\rm (F)}}_{cs}(\vec{\tau},\vec{x})$, and it is defined as
\begin{align}
	D_{cs}(\vec{\tau},\vec{x}) = D_s (\vec{\tau},\vec{x}) {D^{\rm (F)}}_{cs} (\vec{\tau},\vec{x})
\end{align}  
Note also that with the aforementioned decomposition, the \refEq{eq:Observed_distorted_step2} is analysed to 
\begin{align}
	O(\vec{\tau}, \vec{x}) 
	= O_{\rm u}(\vec{\tau}_i,\vec{x}_b) \sum_{s=1}^{N_{\rm s}} D_s(\vec{\tau},\vec{x}) \left\{  \left[1- \sum_{c=1}^{N_{{\rm c} s}} f_{cs}(\vec{\tau},\vec{x}) \right] + 
	\sum_{c=1}^{N_{{\rm c} s}} |\gamma_{cs}  (\vec{\tau}) |^{D_x} f_{cs} (\vec{\tau},\vec{x}) {D^{\rm (F)}}_{cs}(\vec{\tau},  \vec{x}) \right\}
	\; .
\end{align}
This means that we can define the decomposition factor of contamination and decomposition in spacetime functional, namely $\mathcal{DF}(\vec{\tau x})$, as 
\begin{align}
	\mathcal{DF}(\vec{\tau x}) \equiv \sum_{s=1}^{N_{\rm s}} D_s(\vec{\tau},\vec{x}) \left\{  \left[1- \sum_{c=1}^{N_{{\rm c} s}} f_{cs}(\vec{\tau},\vec{x}) \right] + 
	\sum_{c=1}^{N_{{\rm c} s}} |\gamma_{cs}  (\vec{\tau}) |^{D_x} f_{cs} (\vec{\tau},\vec{x}) {D^{\rm (F)}}_{cs}(\vec{\tau},  \vec{x}) \right\}
	\; .
\end{align}
This means that the observable can be defined as 
\begin{align}\label{eq:N_s_objects_N_c_t_contaminants}
	O(\vec{\tau}, \vec{x}) 
	= O_{\rm u}(\vec{\tau}_i,\vec{x}_b) \mathcal{DF}(\vec{\tau x}) 
	\; .
\end{align}
Note that by substituting \refEq{eq:N_s_objects_N_c_t_contaminants} to Eqs. \ref{eq:NPCF} and \ref{eq:NPCFF}, we can calculate the auto- and cross- NPCF and its Fourier transform of $N_{\rm s}$ types of objects, with $N_{{\rm c}t}$ types of contaminants.

In this case, by substituting \refEq{eq:N_s_objects_N_c_t_contaminants} to Eqs. \ref{eq:NPCF}, we get
\begin{align}
	F^{(N)}_{\rm simplified, 3}(\vec{\tau}, \vec{x}, \dots, \vec{x}_{N-1}) = 	O^N_{\rm u}(\vec{\tau}_i, \vec{x}_b)  \langle \mathcal{DF}(\vec{\tau},\vec{s}) \cdot{} 	\mathcal{DF}(\vec{\tau}, \vec{x}+\vec{s}) \cdot{} \dots{} \cdot{} \mathcal{DF}(\vec{\tau}, \vec{x}_{N-1}+\vec{s}) \rangle_s \; ,
\end{align}
In the case where the decomposition function depends only on time, i.e. $\mathcal{DF}(\vec{\tau},\vec{x}) \rightarrow \mathcal{DF}(\vec{\tau})$, we simply get
\begin{align}	
	F^{(N)}_{\rm simplified, 4}(\vec{\tau}, \vec{x}, \dots, \vec{x}_{N-1}) = O^N_{\rm u}(\vec{\tau}_i, \vec{x}_b) \mathcal{DF}^N(\vec{\tau}) \; .
\end{align}

\section{Results} 
We apply our methodology to a variaety of natural systems, i.e. the astronomical scales and the observational scales.

\subsection{Application to a variety of natural systems}
In nature, the most useful summary statistics are the number density fields of a type of objects. The number density fields are usually summarise the number of galaxies and temperature observed in astronomical scales, and also the number of elementary particles in quantum scales. We can call the set of all natural scales, as cosmological scales which include both the astronomical scales as well as the quantum scales.

\subsubsection{Astronomical scales (AS)}
In astronomical scales (AS), we usually use the fluctuations of the number density field of a tracer, with $N_s$ number density of particles, which can be denoted as $\delta_t(\vec{\tau}, \vec{x})=N_s(\vec{\tau}, \vec{x})/\bar{n}_t(\vec{\tau})-1$, where $\bar{n}_t(\vec{\tau})$ is the mean number density of particles of the tracer. We can define that observed LSS (OLSS) tracers belong to the set $S^{\rm OLSS}$, which has $N_{\rm s}^{\rm OLSS}$ such tracers.

The observed matter tracer fluctuation field from $N_{\rm s}^{\rm OLSS}$ tracers is given by
\begin{equation}
		\delta_{\rm O}(\vec{\tau}, \vec{x}) 
		\equiv \sum_{s=1}^{N_{\rm s}^{\rm OLSS}} \delta_{s}(\vec{\tau}, \vec{x}) 
		 \; .
\end{equation}
where $\delta_{m}(\vec{\tau}_i, \vec{x})$ is the matter density fluctuation field, at an initial time $\vec{\tau}_i$.
Ergodic theorem suggest that the hyper-symmetric NPCF of OLSS tracers will be given by
\begin{align}
\xi^{(N)}_{\rm O}(\vec{\tau}, \vec{x}_1, \dots, \vec{x}_{n-1}) &\equiv \langle \delta_{\rm O} (\vec{\tau}, \vec{s}) \cdot{} \delta_{\rm O} (\vec{\tau}, \vec{s}+\vec{x}_1) \cdot{} {\dots} \cdot{} \delta_{\rm O} (\vec{\tau},\vec{s}+\vec{x}_{N-1})  \rangle_s \; .
\end{align}
Note that this formalism includes naturally auto- and cross- correlations between different tracers.
In spatial Fourier space
the observed density field is 
\begin{align}
\delta_{\rm O}(\vec{\tau},\vec{k}) \equiv \int_{\mathcal{M}^{D_x}} d^{D_x} \vec{s}\; e^{-i\vec{k}\cdot{\vec{s}}} \delta_{\rm O}(\vec{\tau},\vec{s}) 
\end{align}
while the N-order correlation function is
\begin{equation}
P^{(N)}(\vec{\tau}, \vec{k}_1, \dots, \vec{ k}_{N-1}) = \langle \delta_{\rm O} (\vec{\tau}, \vec{q}+\vec{k}_1) \cdot{} {\dots} \cdot{} \delta_{\rm O} (\vec{\tau},\vec{q}+\vec{k}_{N-1})  \rangle_q \; ,
\end{equation}
where $\langle \dots \rangle_q \equiv V_{D_q}^{-1} \int_{\mathcal{M}^{D_q}} d^3 \vec{q} $ is the averaged integration of the total observed matter tracer fluctuation field of the OLSS in spatial Fourier space.

Note that this treatment of correlators is a generalisation of study done in \cite{2021arXiv210610278P}, since we now consider a number of tracers of the matter density field, their auto- and cross- correlations and the existence of $D_\tau$ dimensions of conformal times\footnote{Note that \cite{2021arXiv210610278P} has a typo in Eq. $2$, in which the infinitesimal element should be $d^Dx$ and not $d^Nx$. }.

At AS, the main tracers are the ones from the large scale structure (LSS), composed by a variety of $N_s^{\rm LSS}$ different galaxy field types and LSS structures, including Constant MASS galaxies (CMASS), Luminous Red Galaxies (LRG) Emission Line Galaxies (ELG), Quasi Stellar Objects (QSO), Lyman-$\alpha$ lines and their forests (Ly$\alpha$),  (see \cite{2017AJ....154...28B} and references therein), as well as the Intergalactic Medium (IGM) \cite{2021arXiv210810870G}, neutrinos ($\nu$), supermassive black holes (SMBH), gravitational wave sources \cite{2020arXiv200713791B}. This set can be denoted as 
\begin{equation}\label{eq:SetLSS}
S_{\rm LSS} \equiv \left\{\rm  CMASS, LRG, ELG, QSO, Ly\alpha, IGM, \nu, SMBH, GW\ sources, \dots \right\} \; ,
\end{equation}
Note that each tracer can be pixelised in the sky as point sources
\footnote{
Note that possibly, at present, we do detect only tens of neutrinos from extragalactic sources. However, our methodology describes a method that can be applicable in the future in which we do have a better technology to detect more neutrinos from extragalactic sources. As technology evolves and develops, it does mean that we will see more and more of these sources of information, including neutrino from extragalactic sources, and if they are not detectable on earth, they might be detectable by detectors from outerspace.
}.
 This is true for most of the tracers, but yet to be improved by observations regarding the SMBH and GW sources. 

At AS, we also have the matter tracer from the cosmic microwave background (CMB)~\cite{2006PhR...429....1L} and the cosmic infrared background (CIB)~\cite{2015PhRvD..92d3005S}. These can be merged to the cosmic microwave and infrared background CMIB, which define a matter source tracer fluctuation as $\delta_s^{\rm CMIB}$, which takes objects from the set of $S_{\rm CMIB}\equiv S_{\rm CMB} \cup S_{\rm CIB}$. The CMIB is composed by $N^{\rm CMIB}_{\rm s}$ different temperature field types, denoted by the set 
\begin{equation}
S_{\rm CMIB} \equiv \left\{\rm T_{CMB}, E_{CMB},  B_{CMB},  T_{CIB},  E_{CIB},   B_{CIB}, \dots \right\} \; ,
\end{equation}
where T is the temperature fluctuations fields, while E denotes E-polarisation fields, and B denotes B-polarisation fields, of the CMIB (see \cite{2021arXiv210608346I} and references therein). 
Then the OLSS tracers set, can be defined as 
\begin{equation}
	S_{\rm OLSS} \supset S_{\rm LSS} \cup S_{\rm CMIB} \; .
\end{equation}
This means that we are going to have $N_{s}^{\rm OLSS} = N_{\rm s}^{\rm LSS}+N_{\rm s}^{\rm CMIB}$ number of tracers in total.

In the case which the density fluctuation of each source tracer is given by
\begin{align}
	\delta_{s}(\vec{\tau}, \vec{r}) =\delta_m(\vec{\tau}_i,\vec{r}) \sum_{s=1}^{N_{\rm s}^{\rm OLSS}} 
	b_s(\vec{\tau},\vec{r})
	D_s(\vec{\tau},\vec{r})
\end{align}
where $\delta_{m}(\vec{\tau}_i, \vec{x})$ is the matter density fluctuation field, at an initial time $\vec{\tau}_i$, while $b_s(\vec{\tau},\vec{x})$ and $D_s(\vec{\tau},\vec{x})$ are the bias and growth of structure of each tracer. Note that this formalism was inherented by LSS, but it can be easily used in CMIB formalism, since $b_s(\vec{\tau},\vec{x})$ can denote the bias of any tracer of CMIB temperature fluctuations in respect of the matter density fluctuations, and $D_s(\vec{\tau},\vec{x})$ can denote the growth of structure of any CMIB temperature fluctuations. One can use the harmonic decomposition to further simplify the calculation as conceptualised in \cite{2021arXiv210610278P}, but it is beyond the scope of our study.
Using the aforementioned formalism, we have
\begin{align*}
\xi^{(N)}_{\rm O}(\vec{\tau}, \vec{r}_1, \dots, \vec{r}_{N-1}) \equiv 
\langle 
&\sum_{s=1}^{N_{\rm s}^{\rm OLSS}} b_s (\vec{\tau}, \vec{s})  D_s (\vec{\tau}, \vec{s}) \\
&\cdot{} 
\sum_{s=1}^{N_{\rm s}^{\rm OLSS}} b_s (\vec{\tau},\vec{s}+\vec{r}_{1}) 
D_s (\vec{\tau},\vec{s}+\vec{r}_{1})  \\
&\cdot{} {\dots} \\
&\cdot{} \sum_{s=1}^{N_{\rm s}^{\rm OLSS}} b_s (\vec{\tau},\vec{s}+\vec{r}_{N-1}) 
D_s (\vec{\tau},\vec{s}+\vec{r}_{N-1})  
\rangle_s \\
&\cdot{} \xi^{(N)}_{\rm m}(\vec{\tau}_i,\vec{r}_1,\dots,\vec{r}_{N-1})
\numberthis
\; .
\end{align*}
where 
\begin{align}
\xi^{(N)}_{\rm m}(\vec{\tau}_i,\vec{r}_1,\dots,\vec{r}_{N-1}) 
&\equiv \langle \delta_{\rm m} (\vec{\tau}_i, \vec{s}) \cdot{} \delta_{\rm m} (\vec{\tau}_i, \vec{s}+\vec{r}_1) \cdot{} {\dots} \cdot{} \delta_{\rm m} (\vec{\tau}_i,\vec{s}+\vec{r}_{N-1})  \rangle_{\vec{s}}
\end{align}
is the NPCF of matter density fluctuations at an initial time $\vec{\tau}_i$. In the case which there are only scale-independent biases and growths of structures for all tracers the latter equation is simplified to
\begin{align}
	\xi^{(N)}_{\rm O}(\vec{\tau}, \vec{r}_1, \dots, \vec{r}_{N-1}) \equiv \xi^{(N)}_{\rm m}(\vec{\tau}_i,\vec{r}_1,\dots,\vec{r}_{N-1})  \cdot{}
\left[ \sum_{s=1}^{N_{\rm s}^{\rm OLSS}} b_s (\vec{\tau})  D_s (\vec{\tau}) \right]^{N} 
\; .
\end{align}
In this case we can define the bias and growth of structure functional as
\begin{align}
\mathcal{BD}(\vec{\tau}) = 	\left[ \sum_{s=1}^{N_{\rm s}^{\rm OLSS}} b_s (\vec{\tau})  D_s (\vec{\tau}) \right]
\end{align}
therefore we have
\begin{align}
	\xi^{(N)}_{\rm O}(\vec{\tau}, \vec{r}_1, \dots, \vec{r}_{N-1}) \equiv \xi^{(N)}_{\rm m}(\vec{\tau}_i,\vec{r}_1,\dots,\vec{r}_{N-1})  \cdot{}
\left[ \mathcal{BD}(\vec{\tau}) \right]^{N} 
\; .
\end{align}
In case we would like to neglect cross-correlations, we have that 
\begin{align}
	\xi^{(N), {\rm No Cross}}_{\rm O}(\vec{\tau}, \vec{r}_1, \dots, \vec{r}_{N-1}) \equiv
	\xi^{(N)}_{\rm m}(\vec{\tau}_i,\vec{r}_1,\dots,\vec{r}_{N-1})  \cdot{} 
\sum_{s=1}^{N_{\rm s}^{\rm OLSS}} \left[ b_s (\vec{\tau})  D_s (\vec{\tau}) \right]^{N} \; .
\end{align}
Similarly for a scale-independent bias and growth of structure for each tracer, the NPCF Power spectrum is
\begin{align}
		P^{(N)}_{\rm O}(\vec{\tau}, \vec{k}_1, \dots, \vec{k}_{N-1}) \equiv 
P^{(N)}_{\rm m}(\vec{\tau}_i,\vec{k}_1,\dots,\vec{k}_{N-1}) \cdot{} 
\left[ \sum_{s=1}^{N_{\rm s}^{\rm OLSS}} b_s (\vec{\tau})  D_s (\vec{\tau}) \right]^{N}  \; , 
\end{align}
and 
\begin{align}
	P^{(N), {\rm No Cross}}_{\rm O}(\vec{\tau}, \vec{k}_1, \dots, \vec{k}_{N-1}) \equiv
	P^{(N)}_{\rm m}(\vec{\tau}_i,\vec{k}_1,\dots,\vec{k}_{N-1})  \cdot{} 
\sum_{s=1}^{N_{\rm s}^{\rm OLSS}} \left[ b_s (\vec{\tau})  D_s (\vec{\tau}) \right]^{N} \; .
\end{align}
The $\mathcal{BD}(\vec{\tau})$ functional is important for observations in AS, since its form will affect model selection and measurements of the standard model of cosmology.

\subsubsection{Contaminants in AS tracers}
In the case which the density fluctuation of each tracer has also $N_{{\rm c}t} $ contaminants, then the density fluctuations is given by
\begin{align}
	\delta_O(\vec{\tau}, \vec{r}) 
	&=\sum_{s=1}^{N_{\rm s}^{\rm OLSS}} 
	\left\{ 
	\left[1-  \sum_{c=1}^{N_{{\rm c} s}}f_{cs}(\vec{\tau},\vec{x}) \right]\delta_{s}(\vec{\tau},\vec{r})
	+ 
	 \sum_{c=1}^{N_{{\rm c} s}}
	f_{cs}(\vec{\tau},\vec{x}) \delta_{c t}(\vec{\tau}, \vec{\gamma_{cs}}\hspace{1mm}^{-1} \cdot{} \vec{r} ) \right\} \\
\delta_O(\vec{\tau}, \vec{r}) 
&= \delta_m(\vec{\tau}_i, \vec{r})  \; 
\mathcal{FBD}(\vec{\tau},\vec{x}) 
	\; ,
\end{align}
where we have defined the factor contaminant, bias, growth of structure functional with the symbol, 
\begin{align}
\mathcal{FBD}(\vec{\tau},\vec{x}) 
= \sum_{s=1}^{N_{\rm s}^{\rm OLSS}}
\left\{
\left[1- \sum_{c=1}^{N_{{\rm c} s}}f_{cs}(\vec{\tau},\vec{x}) \right]
b_s(\vec{\tau},\vec{x})
D_s(\vec{\tau},\vec{x}) 
+ 
 \sum_{c=1}^{N_{{\rm c} s}}
|\vec{\gamma}_{cs}|^{D_{x}}(\vec{\tau},\vec{x}) f_{cs}(\vec{\tau},\vec{x})
b_{cs}(\vec{\tau},\vec{x})
D_{cs}(\vec{\tau},\vec{x}) \right\}
\end{align}

Note that this means that physically we have that
\begin{align}
	 \mathcal{BD}(\vec{\tau},\vec{x}) = \lim_{ \left\{ f_{cs} (\vec{\tau},\vec{x}) \rightarrow 0 , \forall cs \in S_{cs} \right\}  } \left\{ \mathcal{FBD}(\vec{\tau},\vec{x}) \right\}
\end{align}
while it also means that 
\begin{align}
	 \mathcal{FD}(\vec{\tau},\vec{x}) = \lim_{ \left\{ b_{cs} (\vec{\tau},\vec{x}) \rightarrow 1 , \forall cs \in S_{cs} \right\}  } \left\{ \mathcal{FBD}(\vec{\tau},\vec{x}) \right\}
\end{align}
and 
\begin{align}
	 \mathcal{D}(\vec{\tau},\vec{x}) = \lim_{ \left\{ f_{cs} (\vec{\tau},\vec{x}) \rightarrow 0 , \forall cs \in S_{cs} \right\}  } \lim_{ \left\{ b_{cs} (\vec{\tau},\vec{x}) \rightarrow 1 , \forall cs \in S_{cs} \right\}  } \left\{ \mathcal{FBD}(\vec{\tau},\vec{x}) \right\}
\end{align}
In the case which there are only scale-independent contaminanant factors, biases and growths of structures for all tracers and contaminants, i.e. $\mathcal{FBD}(\vec{\tau},\vec{x}) \rightarrow \mathcal{FBD} (\vec{\tau})$, we have that the NPCF is
\begin{align}
	\frac{\xi^{(N)}_{\rm O}(\vec{\tau}, \vec{r}_1, \dots, \vec{r}_{N-1})}{\xi^{(N)}_{\rm m}(\vec{\tau}_i,\vec{r}_1,\dots,\vec{r}_{N-1}) } 
	\equiv \left\{
\mathcal{FBD}(\vec{\tau}) 
\right\}^{N} 
\; ,
\end{align}
while in fourier space we have
\begin{align}
	\frac{P^{(N)}_{\rm O}(\vec{\tau}, \vec{k}_1, \dots, \vec{k}_{N-1})}{P^{(N)}_{\rm m}(\vec{\tau}_i,\vec{k}_1,\dots,\vec{k}_{N-1}) } 
	\equiv \left\{
\mathcal{FBD}(\vec{\tau}) 
\right\}^{N} 
\; .
\end{align}
The $\mathcal{FBD}(\vec{\tau})$ functional is important for observations in large range of the AS, since its form will affect model selection and measurements of the standard model of cosmology.

\subsubsection{Simplified $N$-point correlators for AS}
The generalised $N$-correlators are difficult to be computed, and therefore until now $N \leq 3$ were extensively used in the literature for $(D_\tau,D_x)=(1,3)$. In this case the observers use the redshift, $z$, as a measure of time, and the three dimensional space for measuring the density fluctuations of matter tracers. This means that the space is going to be reduced to $(\vec{\tau},\vec{x}) \rightarrow (z,\vec{r})$ or $(\vec{\tau},\vec{k}) \rightarrow (z,\vec{k})$, where the two latter vector denotes three dimensions. Therefore here we are listing the correlators for $N\leq 10$, which are going to be used the next about 5-10 years extensively from astronomers. Note that in this case the distortion factors can be defined as the distortion parameter for the perpendicular and parallel to the line-of-sight
\begin{align}
	\gamma_{cs,\perp} &= \frac{D_A(z_s)}{D_A(z_c)} \\
	\gamma_{cs,||}       &= \frac{(1+z_s)/H(z_s)}{(1+z_c)/H(z_c)} \; ,
\end{align}
where $z_s$ is the target's redshift, $z_c$ is the contaminant's redshift, $H(z)$ is the hubble expansion rate, and $D_A(z)$ is the angular distance. Note that in this case we have
\begin{align}
	|\gamma_{cs}(z)|^3 &= \gamma_{cs,\perp}^2\gamma_{cs,||} \; .
\end{align}

In this section, we assume that the contaminant factor, biases, and growths of structures for all targeted and contaminants are scale independent.
Notice that with the aforementioned simplification 
the bias, growth of structure functional becomes:
\begin{align}
	\mathcal{BD}(z) = 
	\sum_{s=1}^{N_{\rm s}^{\rm OLSS}}
b_s(z)
D_s(z) \; ,
\end{align}
while the factor contaminant, bias, growth of structure functional becomes:
\begin{align}\label{eq:FBDz}
\mathcal{FBD}(z)
&=
\sum_{s=1}^{N_{\rm s}^{\rm OLSS}}
\left\{
\left[1- \sum_{c=1}^{N_{{\rm c} s}}f_{cs}(z) \right]
b_s(z)
D_s(z) 
+ 
 \sum_{c=1}^{N_{{\rm c} s}}
\gamma_{cs,\perp}^2\gamma_{cs,||} f_{cs}(z)
b_{cs}(z)
D_{cs}(z) \right\}	
\end{align}
These mean that the observed-relative-to-the-matter N-point correlator in real space becomes 
\begin{align}
	\frac{\xi^{(N)}_{\rm O}(z, \vec{\tilde{r}}_N )}{\xi^{(N)}_{\rm m}(z,\vec{\tilde{r}}_N) } 
	\equiv \left\{
\mathcal{FBD}(z)
 \right\}^{N} 
\; ,
\end{align}
where $\vec{\tilde{r}}_N = (\vec{r}_1,\dots,\vec{r}_{N-1})$, 
while in Fourier space it becomes
\begin{align}
	\frac{P^{(N)}_{\rm O}(z, \vec{\tilde{k}}_N )}{P^{(N)}_{\rm m}(z,\vec{\tilde{k}}_N) } 
	\equiv \left\{
\mathcal{FBD}(z)
 \right\}^{N} 
\; ,
\end{align}
where $\vec{\tilde{k}}_N = (\vec{k}_1,\dots,\vec{k}_{N-1})$. We have coded up factor contaminant, bias, growth of structure function, with some examples to the \href{https://github.com/lontelis/FBDz/blob/main/README.md}{\texttt{FBDz}} code.


We find that for the current interpretation of the astronomical scales, the problem of model selection can be reduced from a $D_\tau$-dimensional manifold, $\mathcal{M}^{D_\tau}$, to a redshift $(D_\tau)$-dimensional manifold, $\mathcal{M}^{D_z}$, using the observed functional form of the contaminant, bias and growth of structure as a function of redshift, formally written as $\mathcal{FBD}(\vec{\tau}) \rightarrow \mathcal{FBD}(z)$, as a well as the standard NPCF of the matter density field, their input functions and parameter dependences. This means that anything that affects the modelling and observation of the factor contaminant, bias and growth of structure functional, will affect also the model selection, and parameter inferences from current and future cosmological surveys and experiments.

\begin{itemize}
\item \textbf{	$2$-point correlators: correlation function and power spectrum.}
The observed-relative-to-the-matter 2-point correlation function in real space becomes 
\begin{align}
\frac{\xi_{\rm O}(z, \vec{r} )}{\xi_{\rm m}(z,\vec{r}) } 
\equiv
\frac{\xi^{(2)}_{\rm O}(z, \vec{r} )}{\xi^{(2)}_{\rm m}(z,\vec{r}) } 
\equiv \left\{
\mathcal{FBD}(z)
 \right\}^2
\end{align}
while in Fourier space, the power spectrum becomes 
\begin{align}
\frac{P_{\rm O}(z, \vec{k} )}{P_{\rm m}(z,\vec{k}) } = 
\frac{P^{(2)}_{\rm O}(z, \vec{k} )}{P^{(2)}_{\rm m}(z,\vec{k}) } 
	\equiv \left\{
\mathcal{FBD}(z)
 \right\}^{2} 
\end{align}

\item \textbf{	$3$-point correlators: 3pt correlation function and bispectrum.}
The observed-relative-to-the-matter 2-point correlation function in real space becomes 
\begin{align}
\frac{\zeta_{\rm O}(z, \vec{r}_1, \vec{r}_2 )}{\zeta_{\rm m}(z,\vec{r}_1, \vec{r}_2) } 
\equiv
\frac{\xi^{(3)}_{\rm O}(z, \vec{r}_1, \vec{r}_2 )}{\xi^{(3)}_{\rm m}(z,\vec{r}_1, \vec{r}_2) } 
	\equiv \left\{
\mathcal{FBD}(z)
 \right\}^{3} 
\end{align}
while in Fourier space, the bispectrum becomes 
\begin{align}
\frac{B_{\rm O}(z, \vec{\tilde{k}}_3 )}{B_{\rm m}(z,\vec{\tilde{k}}_3) } 
= \frac{P^{(3)}_{\rm O}(z, \vec{\tilde{k}}_3 )}{P^{(3)}_{\rm m}(z,\vec{\tilde{k}}_3) } 
\equiv \left\{
\mathcal{FBD}(z)
 \right\}^{3} 
\end{align}
\item \textbf{	$4$-point correlators: 4pt correlation function and trispectrum.}
The observed-relative-to-the-matter 2-point correlation function in real space becomes 
\begin{align}
\frac{\xi^{(4)}_{\rm O}(z, \vec{\tilde{r}}_4 )}{\xi^{(4)}_{\rm m}(z,\vec{\tilde{r}}_4) } 
	\equiv \left\{
\mathcal{FBD}(z)
 \right\}^{4} 
\end{align}
while in Fourier space, the trispectrum becomes 
\begin{align}
\frac{\mathcal{T}_{\rm O}(z, \vec{\tilde{k}}_4 )}{\mathcal{T}_{\rm m}(z,\vec{\tilde{k}}_4) } 
	\equiv \left\{
\mathcal{FBD}(z)
 \right\}^{4} 
\end{align}
\item \textbf{	$5$-point correlators: 5pt correlation function and quadspectrum.}
The observed-relative-to-the-matter 2-point correlation function in real space becomes 
\begin{align}
\frac{\xi^{(5)}_{\rm O}(z, \vec{\tilde{r}}_5 )}{\xi^{(5)}_{\rm m}(z,\vec{\tilde{r}}_5) } 
	\equiv \left\{
\mathcal{FBD}(z)
 \right\}^{5} 
\end{align}
while in Fourier space, the quadspectrum becomes 
\begin{align}
\frac{\mathcal{Q}_{\rm O}(z, \vec{\tilde{k}}_5 )}{\mathcal{Q}_{\rm m}(z,\vec{\tilde{k}}_5) } 
	\equiv \left\{
\mathcal{FBD}(z)
 \right\}^{5} 
  \; .
\end{align}
\item \textbf{	$10$-point correlators: 10pt correlation function and x-spectrum.}
The observed-relative-to-the-matter 10-point correlation function in real space becomes 
\begin{align}
\frac{\xi^{(10)}_{\rm O}(z, \vec{\tilde{r}}_{10} )}{\xi^{(10)}_{\rm m}(z,\vec{\tilde{r}}_{10}) } 
	\equiv \left\{
\mathcal{FBD}(z)
 \right\}^{10} 
\end{align}
while in Fourier space, the x-spectrum becomes 
\begin{align}
\frac{P^{(10)}_{\rm O}(z, \vec{\tilde{k}}_{10} )}{P^{(10)}_{\rm m}(z,\vec{\tilde{k}}_{10}) } 
	\equiv \left\{
\mathcal{FBD}(z)
 \right\}^{10} 
  \; .
\end{align}

\end{itemize}

    \begin{figure*}[ht!]
    \centering 
    \includegraphics[width=120mm]{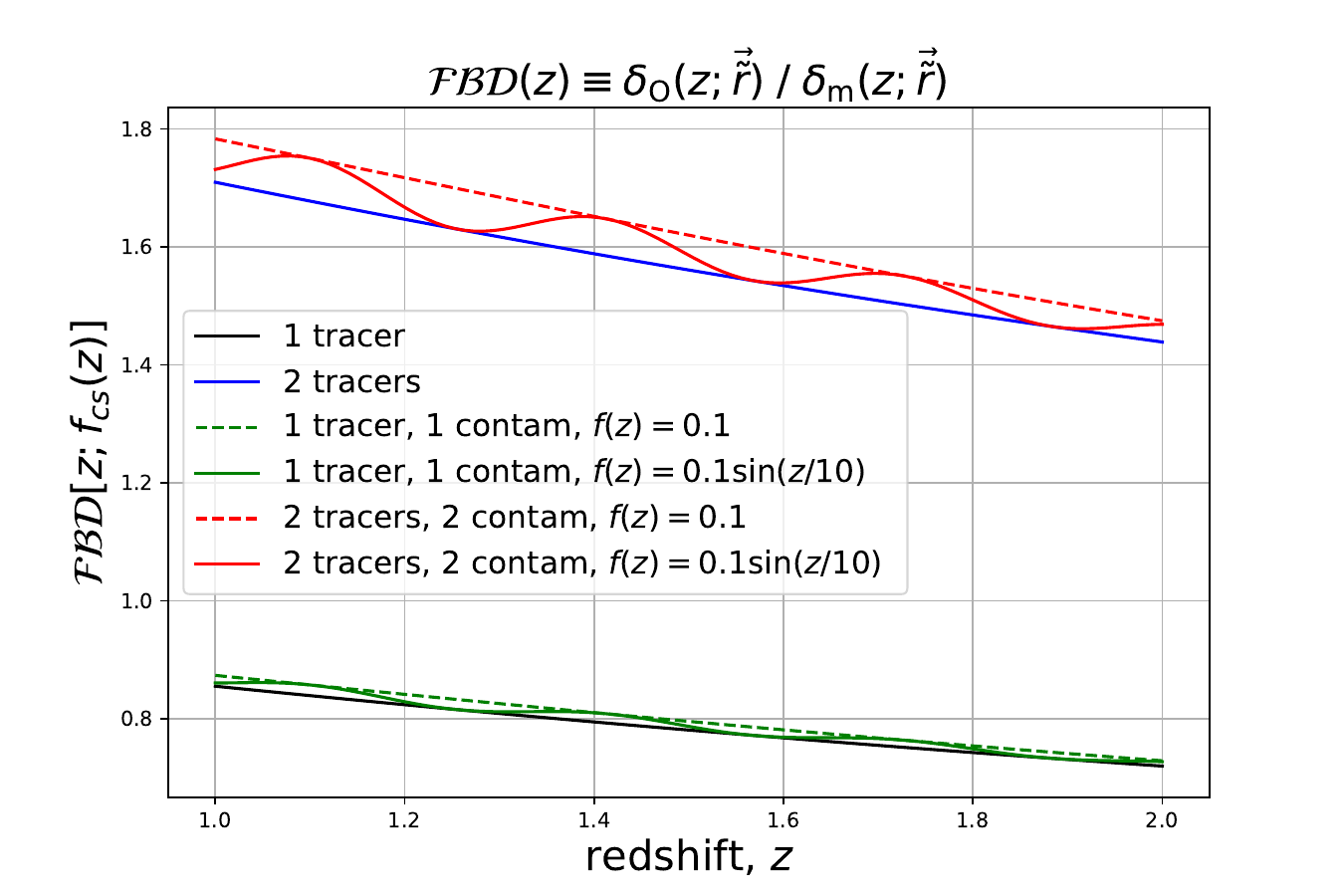} 
    \caption{\label{fig:FBD_N1correlators_example}   Example of the factor contaminant (contam), bias and growth of structure functional, $\mathcal{FBD}$, as a function of redshift, $z$, for $N=1$ order of correlation functional and different flavours of bias, and contaminant factors. The black line represents one tracer, with one bias, and no contaminant factor, while blue line represents two tracers with the same bias model and no contaminant factor. The green (red) dash line represents one (two) tracer(s) contaminated with a constant contaminant factor, $f(z)=10\%$. The same colors but with continuous line represent the same information but for a redshift evolved contaminant factor, $f(z)=0.1\sin(z/10)$.   [See \refS{sec:Application_cosmological_scales}] }
    \end{figure*}

\subsubsection{An application on current concordance cosmology}\label{sec:Application_cosmological_scales}

We assume a simple cosmological model which describes part of the cosmological scales, i.e. the LSS and CMIB scales as follows.
We consider the following fiducial concordance cosmology. We assume the speed of light, $c\simeq 3 \times 10^{8}$ m/s,
the dimensionless Hubble constant, $h=0.67$; the present baryon density ratio, $\Omega_{\rm b, 0}=0.05$; the present matter density ratio, $\Omega_{\rm m, 0}=0.32$; present dark energy density ratio, $\Omega_{\Lambda, 0} =  0.68$; the primordial power spectrum scalar amplitude, $A_{\rm s}=2.1 \times 10^{-9}$; the spectral index, $n_{\rm s}=0.97$. We neglect the neutrino mass, $\sum_{i}m_{\nu_{i}}=0$ eV while the effective number of neutrinos is, $N_{\rm eff}=3.046$. We assume general relativity, by imposing that the growth rate has $\gamma=0.545$ \citep{1991MNRAS.251..128L,Growth}.

    \begin{figure*}[ht!]
    \centering 
    \includegraphics[width=120mm]{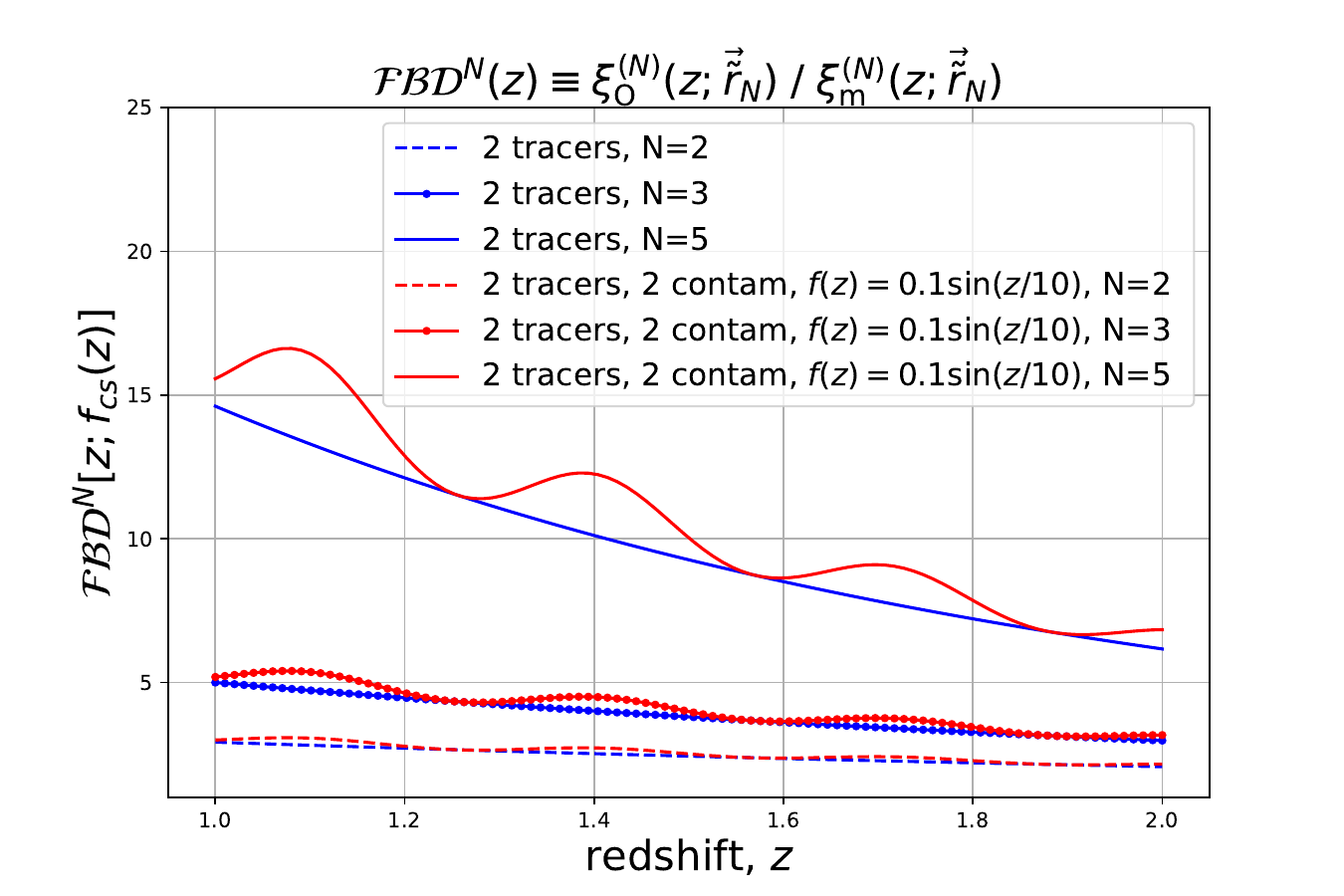} 
    \caption{\label{fig:FBD_Ncorrelators_example}   Example of the factor contaminant, bias and growth of structure functional, $\mathcal{FBD}$, as a function of redshift, $z$, for different $N$ order of correlation functional for two tracers assuming the same bias, and the same redshift evolved contaminant factor, $f(z)=0.1\sin(z/10)$. The blue (red) line represents two tracers with (without) contamination. The dash (dotted or continuous) line represents the order N=2 (3 or 5) of the correlation function. [See \refS{sec:Application_cosmological_scales}] }
    \end{figure*}
    
Note that we can build an example by choosing a particular contaminant factor, bias and growth of structure model.
We choose the following set as follows. We choose a simple model for the growth of structure
\begin{align}
	D(z) = \exp\left\{  \int_0^z \diff z'\, \left[ \frac{\Omega_{\rm m, 0}\, (1+z')^3 }{ H^2(z')/H^2_0 } \right]^{\gamma} \frac{-1}{1+z'} 
	\right\}\; .
\end{align}
We choose two functions which simulate some observations for the contaminant factor
\begin{align}
	f_1(z) &= f_0  \\
	f_2(z) &= f_0 \sin(z/10) \; ,
\end{align}
where $f_0$ is considered a free parameter with fiducial value $10\%$. We choose one popularly observed function for the deterministic bias model as, 
\begin{align}
	b(z) &= b_0 \sqrt{1+z} \; ,
\end{align}
where $b_0$ is a free parameter with fiducial value the unity. 

\begin{figure*}[ht!]
    \centering 
    \includegraphics[width=120mm]{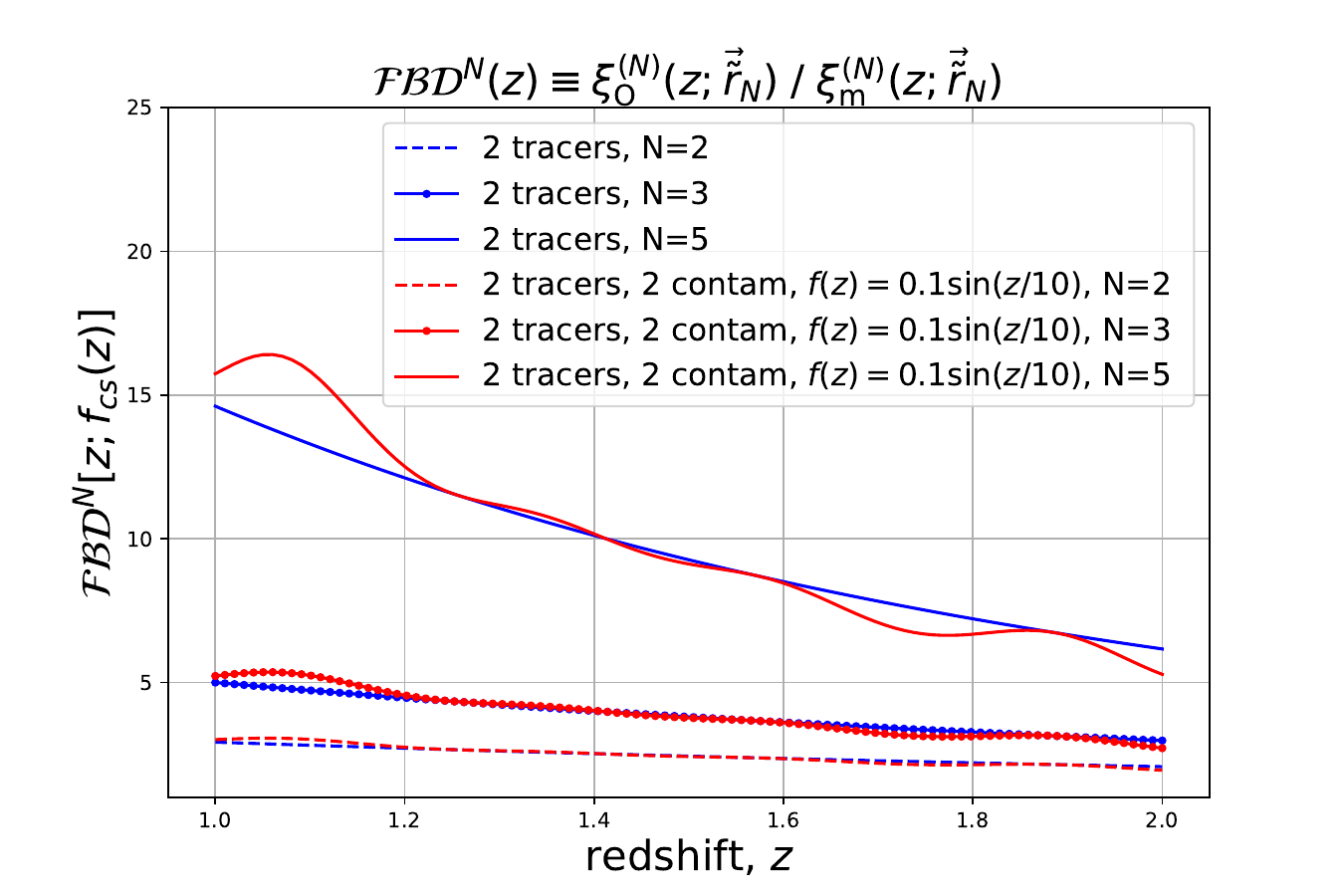} 
    \caption{\label{fig:FBD_Ncorrelators_examplez_contam_smaller} Similar to \refF{fig:FBD_Ncorrelators_example} with the difference that the contaminant redshift regions has smaller redshifts than the targeted one, i.e. the contaminant redshift region is $z_{\rm c} \in \left[0.2,0.8\right]$. [See \refS{sec:Application_cosmological_scales}] }
    \end{figure*}

In our application, we assume a targeted redshift range of interest, $1\leq z\leq 2$, and a contaminant redshift range of interest, $z_{\rm c} \in \left[2.0,2.5\right]$ for the first two examples, while for the last one we assume a contaminant redshift range which has smaller redshift values than the targeted one, i.e. $z_{\rm c} \in \left[0.2,0.8\right]$. 
In Figs. \ref{fig:FBD_N1correlators_example}, \ref{fig:FBD_Ncorrelators_example} and \ref{fig:FBD_Ncorrelators_examplez_contam_smaller}, we present some quantitative examples of the factor contaminant, bias, and growth of structure functional as a function of redshift, $\mathcal{FBD}(z)$, as constructed by \refEq{eq:FBDz}. For all cases in which we apply a sinusodial behaviour for the factor of contaminant, there is a sinusodial effect on the observed functional $\mathcal{FBD}(z)$, which is represented in all aforementioned figures.

From \refF{fig:FBD_N1correlators_example}, we find as expected that 
\begin{enumerate}
	\item increasing (decreasing) number of tracers
	\item increasing (decreasing) contaminant factor
\end{enumerate}
results to an increasing (decreasing) functional, $\mathcal{FBD}(z)$. Doubling the number of tracers, existing in the same redshift region with the same bias model, results to a doubling of $\mathcal{FBD}(z)$. A $10\%$ increase of the contaminant factor results to a 2-1\% ( 4-2.4\%  ) increase of the fuctional, $\mathcal{FBD}(z)$, for one (two) tracer in the targeted redshift range of interest. 

From \refF{fig:FBD_Ncorrelators_example}, we additionally find that increasing (decreasing) order of correlation N, results to an increasing (decreasing) functional, $\mathcal{FBD}(z)$. For two tracers in the redshift range of interest, $1\leq z\leq 2$, an $10\%$ increase of the contaminant factor results to 20-10\% ( 7-5\% ) increase of the functional, $\mathcal{FBD}(z)$, for $N=5 ( 2 )$ order of correlation.

From \refF{fig:FBD_Ncorrelators_examplez_contam_smaller}, we find that for two tracers, for order of correlation $N=5 (2)$, a $10\%$ increase of the contaminant factor from contaminants in lower redshift region, $z_{\rm c} \in [0.2,0.8]$, either produces up to a $18\%$ ($7\%$) increase of the $\mathcal{FBD}(z)$ functional from redshifts lower than $z_s \lesssim 1.4$, or produces up to a $17\%$  ($6\%$) decrease of the $\mathcal{FBD}(z)$ functional at higher redshifts $z_s \gtrsim 1.4$. 

Overall, our results suggest that $10\%$ contamination from lower redshift produces up to a $18\%$ increase of the observed functional, $\mathcal{FBD}(z)$, at low redshifts $z_s\sim 1.4$, while a $17\%$ decrease of $\mathcal{FBD}(z_s \gtrsim 1.4)$, as opposed to $10\%$ contamination from higher redshifts, in which only up to a $20\%$ increase is produced, for $N=5$ order of correlation. This means that a special treatment is needed for these lower redshift contaminants.

\subsubsection{Quantum scales (QS)}
Quantum field theory has a long history with NPCF \cite{Peskin:1995ev}. We know that the natural physical quantum systems are at least described by a Minkowski spacetime, in quantum scales. In this study, we expand this type of description in order to include a generalised Minkowski spacetime in NPCF of quantum mechanical systems. As in astronomical systems, we expect that the in quantum systems, there is also the need of a targeted source quantum system and a contaminant one, which can be caused by elements which we would like not to target or observe. Therefore we can construct an NPCF in a  such an object as we have achieved for astronomical scales.
In quantum field theory an NPCF is described using a quantum field, $\phi(\vec{\tau x})\equiv \phi(\vec{\tau},\vec{x}) $ using the equation
\begin{align}
	C_N \equiv C_N(\vec{\tau}, \vec{x}_1, \dots, \vec{x}_{N}) &= \langle \phi(\vec{\tau}, \vec{x}_1) \dots \phi(\vec{\tau}, \vec{x}_{D_x})  \rangle_\phi \\
	&=  \int_{\mathcal{M}^\phi} \mathcal{D} \phi \; \phi(\vec{\tau}, \vec{x}_1) \dots \phi(\vec{\tau}, \vec{x}_{N}) \; \exp \left\{ {\frac{i}{\hbar} S[\vec{\tau x},\phi(\vec{\tau x}) ] } \right\}
\end{align}
where $S[\vec{\tau x},\phi]$ is the action which describes the physical system, $\hbar$ is the reduced Planck constant, $\mathcal{M}^\phi$ is the manifold of the field $\phi$, and we have used the completeness relation, i.e. $\int_{ \mathcal{M}^\phi } \mathcal{D}\phi \; \exp\left\{ {\frac{i}{\hbar} S[\vec{\tau x},\phi(\vec{\tau x}) ]} \right\}  = 1$.
Note that in case which we would like to distinguish a targeted object category, in respect of several others which contaminate the targeted object category we can think the following, we can apply the description in \refS{sec:distortion_from_contaminants}. Then for some targeted quantum field, $\phi_s(\vec{\tau x})$, some contaminated targeted quantum field, $\phi_{cs}(\vec{\tau x})$ and their respective decomposition functionals $D_s^{\phi}(\vec{\tau x})$, $D_{cs}^\phi(\vec{\tau x})\equiv D_s^{\phi}(\vec{\tau x}) D_{cs}^{(F)\phi}(\vec{\tau x})$ and their universal quantum field functional $\phi_{\rm u} \equiv \phi_u(\vec{\tau_i x_b})$ we have
\begin{align}
	\phi(\vec{\tau x}) 
	= \phi_{\rm u}(\vec{\tau_i x}_b) \sum_{s=1}^{N_{\rm s}} D_{s}^\phi(\vec{\tau},\vec{x}) \left\{  \left[1- \sum_{c=1}^{N_{{\rm c} s}} f_{cs}(\vec{\tau},\vec{x}) \right] + 
	\sum_{c=1}^{N_{{\rm c} s}} |\gamma_{cs}|^{D_x}(\vec{\tau}) f_{cs} (\vec{\tau},\vec{x}) D^{\rm (F) \phi}_{cs}(\vec{\tau},  \vec{x}) \right\}
	\; ,
\end{align}
where $f_{cs} (\vec{\tau},\vec{x})$ is defined as the contaminant factor of targeted elementary particles which are contaminated by any natural contaminants appear at the level of their detections, such as non-targeted elementary particles, from other generalised spacetime regions, or composition of elementary particles, of even cosmic rays. Then,  we can define the contaminated targeted functional as 
\begin{align}
	 \mathcal{C}^{\rm TF}(\vec{\tau x}) = \sum_{s=1}^{N_{\rm s}} D_{s}^\phi(\vec{\tau},\vec{x}) \left\{  \left[1- \sum_{c=1}^{N_{{\rm c} s}} f_{cs} (\vec{\tau},\vec{x}) \right] + 
	\sum_{c=1}^{N_{{\rm c} s}} |\gamma_{cs}|^{D_x}(\vec{\tau}) f_{cs} (\vec{\tau},\vec{x}) D^{\rm (F) \phi}_{cs}(\vec{\tau},  \vec{x}) \right\} \; .
\end{align}
This means that 
$
	\phi(\vec{\tau x}) 
	= \phi_{\rm u} \;  \mathcal{C}^{\rm TF}(\vec{\tau x}) \; .
$
In this case we have that the NPCF for a quantum field, $\phi$ is analysed as
\begin{align}
	C_N(\vec{\tau}, \vec{x}_1, \dots, \vec{x}_{N}) =
	\int_{\mathcal{M}^\phi_{\rm u}} \mathcal{D}\phi_{\rm u} \left( \phi_{\rm u} \right)^N \int_{\mathcal{M}^{\mathcal{C}^{\rm TF}}}\mathcal{D} \mathcal{C}^{\rm TF} \;  \; \mathcal{C}^{\rm TF}(\vec{\tau}, \vec{x}_1) \dots \mathcal{C}^{\rm TF}(\vec{\tau}, \vec{x}_{N}) \; e^{ \left\{ {\frac{i}{\hbar} S[\vec{\tau x},\phi_{\rm u},\mathcal{C}^{\rm TF}  ] } \right\} }
\end{align}
Note that the use of $C^{\rm FT}(\vec{\tau x})$ functional is the best way to specialise an observed NPCF for a generic quantum field. This formalism can be used in quantum field theory experiments, such as the LHC.

Furthermore, this formalism finds practical utility in analyzing data from particle colliders like the LHC, where NPCFs are routinely computed to probe interactions beyond the Standard Model, such as in searches for supersymmetric particles or dark matter candidates \cite{CMS:2024sus,ATLAS:2018dm}. For instance, consider a scenario involving the decay chains of gluinos in proton-proton collisions, where the targeted quantum field \(\phi_s\) represents the signal process (e.g., \(\tilde{g} \to q\bar{q}\tilde{\chi}^0_1\)), while contaminants \(\phi_{cs}\) arise from background processes like top-quark pair production or QCD jets misidentified due to detector resolution. By incorporating the distortion factor \(\gamma_{cs}\) to account for momentum rescaling from extra-dimensional effects or instrumental smearing, the contaminated NPCF \(C_N\) can quantify deviations from expected Gaussian statistics, enhancing signal extraction in multivariate analyses \cite{ATLAS:2020jets}. This not only mitigates systematic uncertainties but also allows for testing theories with extra dimensions, such as large extra dimensions models \cite{ATLAS:2016gravity}, directly against experimental observables like jet multiplicities or missing transverse energy distributions.

\section{Conclusion}

In this paper, we constructed a mathematical formalism for $(D_\tau,D_x)$-dimensional manifolds with $N$-correlators, i.e. the N-point correlation functional (NPCF) of $N_s$ types of objects with and without cross correlations and/or contaminants. In particular, we build this formalism using simple notions of mathematical physics, field theory, topology, algebra, statistics, N-correlators and Fourier transform. We discuss this formalism in the context of cosmological scales, i.e. from astronomical scales to quantum scales. 

We present and discuss the applicability of this formalism in the context of cosmological scales, i.e. from astronomical scales to quantum scales, for which we give some intuitive examples, explicitly, for standard  spacetime dimensions and extra dimensions. We conclude that this study can be used as a guide to analyse and build models for future cosmological and collider experiments. Furthermore, this study opens the road of extra dimension studies.

We find that for the current interpretation of the astronomical scales, the problem of model selection can be reduced from a $(D_\tau,D_x)$-dimensional manifold, to a (redshift,spatial)-dimensional manifold, $\mathcal{M}^{(D_z,D_x)}$, using the observed functional form of the scale independent contaminant, bias and growth of structure as a function of redshift, formally written as $\mathcal{FBD}(\vec{\tau},\vec{x}) \rightarrow \mathcal{FBD}(z)$, as well as the standard NPCF of the matter density field, their input functions and parameter dependences.
Using current concordance cosmology, a quantitative analysis of a special configuration shows that there is up to a $20\%$ increase of the observed functional, $\mathcal{FBD}(z)$, for two possible matter tracers, in a targeted redshift region of $1\leq  z_s \leq 2$, with two possible contaminants, from higher redshifts, $2 \leq z_{\rm c} \leq 2.5$, for $N\leq 5$ order of correlators. However, there is a dependence of the number of tracers and the redshift direction of these contaminants (lower or higher redshifts). This means that a special treatment is needed for these applications. We conclude that anything that affects the modelling and observation of the factor contaminant, bias and growth of structure functional, will affect also the model selection, and parameter inferences from current and future cosmological surveys and experiments.
Furthermore, in quantum scales, we have found that this formalism corresponds to a specialisation of the NPCF for a generic quantum field used so far. In general, we conclude that this formalism can be used to any current and future cosmological survey and experiment, for model selection and parameter quantification inferences. 

\vspace{1cm}
\hspace{6cm} $\mathcal{O}. \mathcal{E}. \Delta$.

\section*{AKNOWLEDGEMENTS}


PN would like to thank the anonymous reviewer for their insightful and constructive comments, which significantly improved the clarity and presentation of this manuscript. We acknowledge open libraries support \texttt{IPython} \citep{4160251}, \texttt{Matplotlib} \citep{Hunter:2007}, \texttt{NUMPY} \cite{Walt:2011:NAS:1957373.1957466}, \texttt{SciPy 1.0} \citep{2019arXiv190710121V}, \texttt{IMINUIT} \citep{James:1975dr}, 
\href{https://github.com/lontelis/cosmopit}{\texttt{COSMOPIT}}\citep{2017JCAP...06..019N,2018JCAP...12..014N}. 
The \href{https://github.com/lontelis/FBDz/blob/main/README.md}{\texttt{FBDz}} code product of this analysis is publicly available.  

\label{Bibliography}


\bibliographystyle{unsrtnat_arxiv} 

\bibliography{Bibliography} 

\appendix
\section{Statements \& declarations}
\subsection{Funding}
The authors declare that no funds, grants, or other support were received during the preparation of this manuscript.
\subsection{Competing and conflict of interests}
The authors have no relevant financial, non-financial or confict of interests to disclose.
\subsection{Author contributions}
All authors contributed to the study conception and design. Material preparation, data collection and analysis were performed by Pierros Ntelis. The first draft of the manuscript was written by Pierros Ntelis and all authors commented on previous versions of the manuscript. All authors read and approved the final manuscript.
\subsection{Data availability}
Data sharing is not applicable to this article as no new data were created or analyzed in this study. 

\end{document}